\documentclass[aps,prl,twocolumn,superscriptaddress]{revtex4-1}

\usepackage{amsmath}
\usepackage{graphicx}
\usepackage[usenames,dvipsnames]{xcolor}
\usepackage[bookmarks=false,colorlinks]{hyperref}
\hypersetup{
	linkcolor=MidnightBlue,        % color of internal links
	citecolor=blue,     % color of links to bibliography
	filecolor=Plum,      	% color of file links
	urlcolor=MidnightBlue,           % color of external links
}

\usepackage{ccaption}
\usepackage{ragged2e}
  \captionnamefont{\small \bfseries }
  \captiontitlefont{\small }
  \captionstyle{\justifying}

\begin{document}

\title{Quasiparticle Relaxation Dynamics in URu$_{2-x}$Fe$_{x}$Si$_{2}$ Single Crystals}
\author{Peter Kissin}
\affiliation{Department of Physics, University of California San Diego, 9500 Gilman Drive, La Jolla, California 92093, USA}

\author{Sheng Ran}
\altaffiliation[Present Addresses: ]{Center for Nanophysics and Advanced Materials, Department of Physics, University of Maryland, College Park,
MD 20742; NIST Center for Neutron Research, National Institute
of Standards and Technology, 100 Bureau Drive, Gaithersburg, MD
20899.}
\affiliation{Department of Physics, University of California San Diego, 9500 Gilman Drive, La Jolla, California 92093, USA}
\affiliation{Center for Advanced Nanoscience, University of California San Diego, La Jolla, California 92093, USA}

\author{Dylan Lovinger}
\affiliation{Department of Physics, University of California San Diego, 9500 Gilman Drive, La Jolla, California 92093, USA}

\author{Verner K. Thorsm\o lle}
\affiliation{Department of Physics, University of California San Diego, 9500 Gilman Drive, La Jolla, California 92093, USA}

\author{Noravee Kanchanavatee}
\altaffiliation[Present Address: ]{Department of Physics, Chulalongkorn University, Pathumwan, 10330, Thailand.}
\affiliation{Department of Physics, University of California San Diego, 9500 Gilman Drive, La Jolla, California 92093, USA}
\affiliation{Center for Advanced Nanoscience, University of California San Diego, La Jolla, California 92093, USA}

\author{Kevin Huang}
\altaffiliation[Present Address: ]{National High Magnetic Field Laboratory, Florida State University, Tallahassee, FL 32313.}
\affiliation{Center for Advanced Nanoscience, University of California San Diego, La Jolla, California 92093, USA}
\affiliation{Materials Science and Engineering Program, University of California San Diego, 9500 Gilman Drive, La Jolla, California 92093, USA}

\author{M. Brian Maple}
\affiliation{Department of Physics, University of California San Diego, 9500 Gilman Drive, La Jolla, California 92093, USA}
\affiliation{Center for Advanced Nanoscience, University of California San Diego, La Jolla, California 92093, USA}

\author{Richard D. Averitt}
\email[Corresponding Author:]{raveritt@ucsd.edu}
\affiliation{Department of Physics, University of California San Diego, 9500 Gilman Drive, La Jolla, California 92093, USA}

\date{\today}

\begin{abstract}
We investigate quasiparticle relaxation dynamics in URu$_{2-x}$Fe$_{x}$Si$_{2}$ single crystals using ultrafast optical-pump optical-probe (OPOP) spectroscopy as a function of temperature ($T$) and Fe substitution ($x$), crossing from the hidden order (HO) phase ($x$ = 0) to the large moment antiferromagnet (LMAFM) phase ($x$ = 0.12). At low $T$, the dynamics for $x$ = 0 and $x$ = 0.12 are consistent with the low energy electronic structure of the HO and LMAFM phases that emerge from the high $T$ paramagnetic (PM) phase. In contrast, for $x$ = 0.1, two transitions occur over a narrow $T$ range (from ~15.5 - 17.5 K). A PM to HO transition occurs at an intermediate $T$ followed by a transition to the LMAFM phase at lower $T$. While the data at low $T$ are consistent with the expected coexistence of LMAFM and HO, the data in the intermediate $T$ phase are not, and instead suggest the possibility of an unexpected coexistence of HO and PM. Additionally, the dynamics in the PM phase reflect the presence of a hybridization gap as well as strongly interacting spin and charge degrees of freedom. OPOP yields insights into meV-scale electrodynamics with sub-Kelvin $T$ resolution, providing a complementary approach to study low energy electronic structure in quantum materials.

\end{abstract}

\maketitle

The metallic actinide compound URu$_{2}$Si$_{2}$, with its many proximal phases, offers a platform to study emergent phenomena in $f$-electron systems poised between localization and itinerancy. In particular, the hidden order (HO) phase, which develops from a strongly correlated paramagnetic (PM) phase below $T_{0}$ = 17.5 K \cite{Palstra1985, Maple1986}, has attracted extensive attention \cite{Mydosh2011}. The combined efforts of ARPES \cite{Meng2013, Bareille2014, Chatterjee2013, Boariu2013}, quantum oscillations \cite{Hassinger2010}, and band structure calculations \cite{Elgazzar2009, Oppeneer2011} have led to a consistent picture of the Fermi surface. Neutron scattering measurements have identified magnetic excitations at $Q_{0} = (1,0,0)$ and $Q_{1} = (1\pm0.4,0,0)$ in the body-centered tetragonal Brillouin zone (BZ) of the PM phase, which are gapped in the HO phase \cite{Wiebe2007, Butch2015}. Despite this progress, the order parameter of the HO phase remains unidentified, motivating novel experimental approaches.

An alternate route to understanding HO is to instead study the large-moment antiferromagnetic (LMAFM) phase in pressurized URu$_{2}$Si$_{2}$ \cite{Motoyama2003}. While the LMAFM and HO phases have similar signatures in thermodynamics and transport \cite{Hassinger2008}, and nearly identical Fermi surfaces \cite{Hassinger2010}, the order parameter and symmetries of the LMAFM phase are known, facilitating progress in theory \cite{Oppeneer2011}. Unfortunately, even the modest pressure necessary to access LMAFM renders many techniques impossible. However, substitution of Fe for Ru yields an antiferromagnetic phase without applied pressure \cite{Kanchanavatee2011, Das2015}. Striking resemblances exist between the magnetic excitation spectra of the two phases \cite{Butch2016}, and distinctive features of the phase diagrams of URu$_{2}$Si$_{2}$ are reproduced \cite{Ran2016, Ran2017}. Apparently, Fe substitution acts as a chemical pressure, enabling new measurements in the LMAFM phase \cite{Kung2016}.

Optical Pump Optical Probe (OPOP) spectroscopy has been used to study quasiparticle (QP) relaxation dynamics in heavy fermion compounds \cite{Demsar2003_2, Demsar2006, Demsar2006_2, Chia2006, Talbayev2010}. The versatility of this technique comes from its extreme sensitivity to the formation of meV-scale gaps in the electronic density of states (DOS) near the Fermi Energy $E_{F}$. The presence of a gap can be inferred from the temperature ($T$) and pump fluence ($F$) dependence of the QP relaxation dynamics and may result in an increase in the relaxation time by several orders of magnitude at low $T$.

\begin{figure} \includegraphics[width=2.375 in,clip]{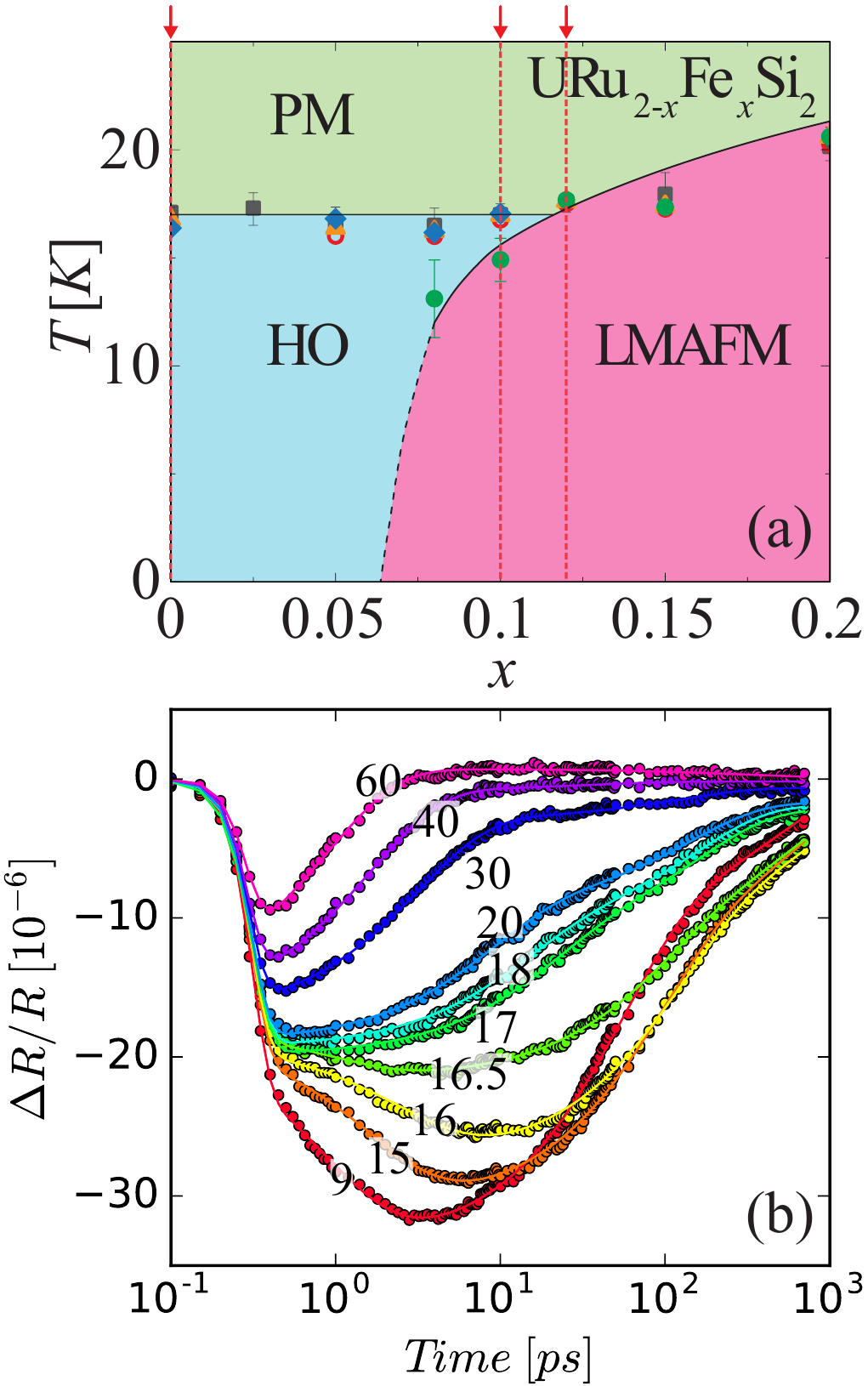}
\caption{(a) Phase diagram of URu$_{2-x}$Fe$_{x}$Si$_{2}$, reproduced from \cite{Ran2016}. Transition temperatures determined by resistivity, magnetization, and heat capacity are depicted by gray squares, red rings, and orange triangles, respectively. Thermal expansion shows two transitions, which are depicted by blue diamonds and green circles. Optical pump-probe data are presented on samples with $x$ = 0, $x$ = 0.1, and $x$ = 0.12, indicated by the red dashed lines and arrows. (b) Fractional change in reflectivity $\Delta$R/R vs. time after photoexcitation for $x$ = 0. Each curve is labelled by the corresponding $T$ in Kelvin. Solid lines are fits to the data using Eqn. \ref{eq:1}. \label{fig:1}}
\end{figure}

In this letter, we investigate QP relaxation dynamics in URu$_{2-x}$Fe$_{x}$Si$_{2}$ single crystals spanning a broad range of Fe substitution ($x$), focusing on the compositions indicated in Fig. \ref{fig:1}(a). We observe differences in the dynamics between the HO ($x$ = 0) and LMAFM ($x$ = 0.12) phases, which are successfully described using a simple phenomenological model of relaxation bottlenecks associated with gaps characteristic of each state. In contrast, for $x$ = 0.1, two transitions occur over a narrow $T$ range (from ~15.5 - 17.5 K). A PM to HO transition occurs at higher $T$, with a subsequent transition to a LMAFM phase at lower $T$. While signatures of heterogeneity are present in both phases, anomalies in the intermediate $T$ HO phase suggest the unusual possibility of a persistent PM volume fraction. In the PM phase, the dynamics reveal the presence of a hybridization gap as well as strongly interacting spin and charge degrees of freedom.

The Fe-substituted single crystals were grown in a tetra-arc furnace using the Czochralski technique \cite{Ran2016}. OPOP measurements used 25 fs laser pulses centered at 800 nm with a repetition rate of 209 kHz \cite{Supp}. The cross polarized pump and probe beams were focused to $\frac{1}{e^{2}}$ spot diameters of 100 $\mu$m and 60 $\mu$m respectively. $F$ was fixed at 0.5 $\mu$J/cm$^{2}$ for all measurements to ensure minimal heating of the sample \cite{Supp}. The data were collected from large, flat areas of samples cleaved in the $a$-$b$ plane and placed in a continuous flow liquid He optical cryostat.

Fig. \ref{fig:1}(b) shows the photoinduced change in fractional reflectivity $\Delta$R/R as a function of time for $x$ = 0 (For other samples, see \cite{Supp}). The dynamics are qualitatively similar for all samples. At high $T$, the relaxation is biexponential, consisting of a fast, negative component with a time constant of hundreds of fs and a small, slow, positive component with a time constant of hundreds of ps. Upon cooling, the fast component begins to slow to the few ps timescale, and the slower component switches sign and increases in amplitude. At $T_{0}$, between 17 K and 16.5 K, the signal amplitude continues to increase for a few ps after photoexcitation \cite{Supp} and the relaxation time approaches a ns. These abrupt changes to the dynamics at $T_{0}$ mark the transition to the low $T$ phase.

We fit the data with a multiexponential function:
\begin{equation}\label{eq:1}
\frac{\Delta R}{R}(t) = f(t)*(A_{f}e^{-t/\tau_{f}}+A_{s}e^{-t/\tau_{s}}+C)
\end{equation}
Where $f(t)=0.5*(1-erf[-\sigma(t-t_{0})])*(1+A_{r}(1-e^{-(t-t_{0})/\tau_{r}}))$. In $f(t)$, the first term containing the error function represents the fast rise present at all $T$. This term is included for completeness and $\sigma$ is $T$ independent. The term containing $A_{r}$ and $\tau_{r}$ represents the slow rise dynamics that onset below $T_{0}$. The second term in Eqn. \ref{eq:1} contains two exponential decays and a constant. The constant is close to the experimental noise floor of $10^{-6}$ at all $T$, so our analysis will focus on the $T$ dependence of the parameters from the exponential terms. In order to compare amplitudes above and below $T_{0}$, we define $(1+A_{r})\*A_{f,s}=A_{1,2}$ ($\tau_{f,s}=\tau_{1,2}$ for consistency).

Relaxation of photoexcited QPs in the presence of a gap requires e-h recombination with the emission of a high energy boson (HEB) with energy h$\omega$ $\geq$ E$_{gap}$. This situation is frequently analyzed using the phenomenological Rothwarf-Taylor (RT) model \cite{Rothwarf1967}. The key parameters in the RT model are the bare QP recombination rate $\gamma_{r}$, the rate of across-gap QP excitation by a HEB $\gamma_{pc}$, and the rate of escape or anharmonic decay of HEBs $\gamma_{esc}$. Various regimes are realized depending on these rates \cite{Kabanov2005, Torchinsky2010}. If $\gamma_{r} \gg \gamma_{pc}$ or $\gamma_{esc} \gg \gamma_{pc}$, then bimolecular recombination dynamics are observed, and the bare recombination rate of QPs $\gamma_{r}$ can be obtained. On the other hand, if $\gamma_{pc}$ is the fastest rate, the result is a strong bottleneck with a relaxation rate limited to $\gamma_{esc}$. In URu$_{2}$Si$_{2}$, the relaxation dynamics are independent of $F$, implying strongly bottlenecked QP relaxation \cite{Supp}.

\begin{figure} \centering \includegraphics[width=3.375 in,clip]{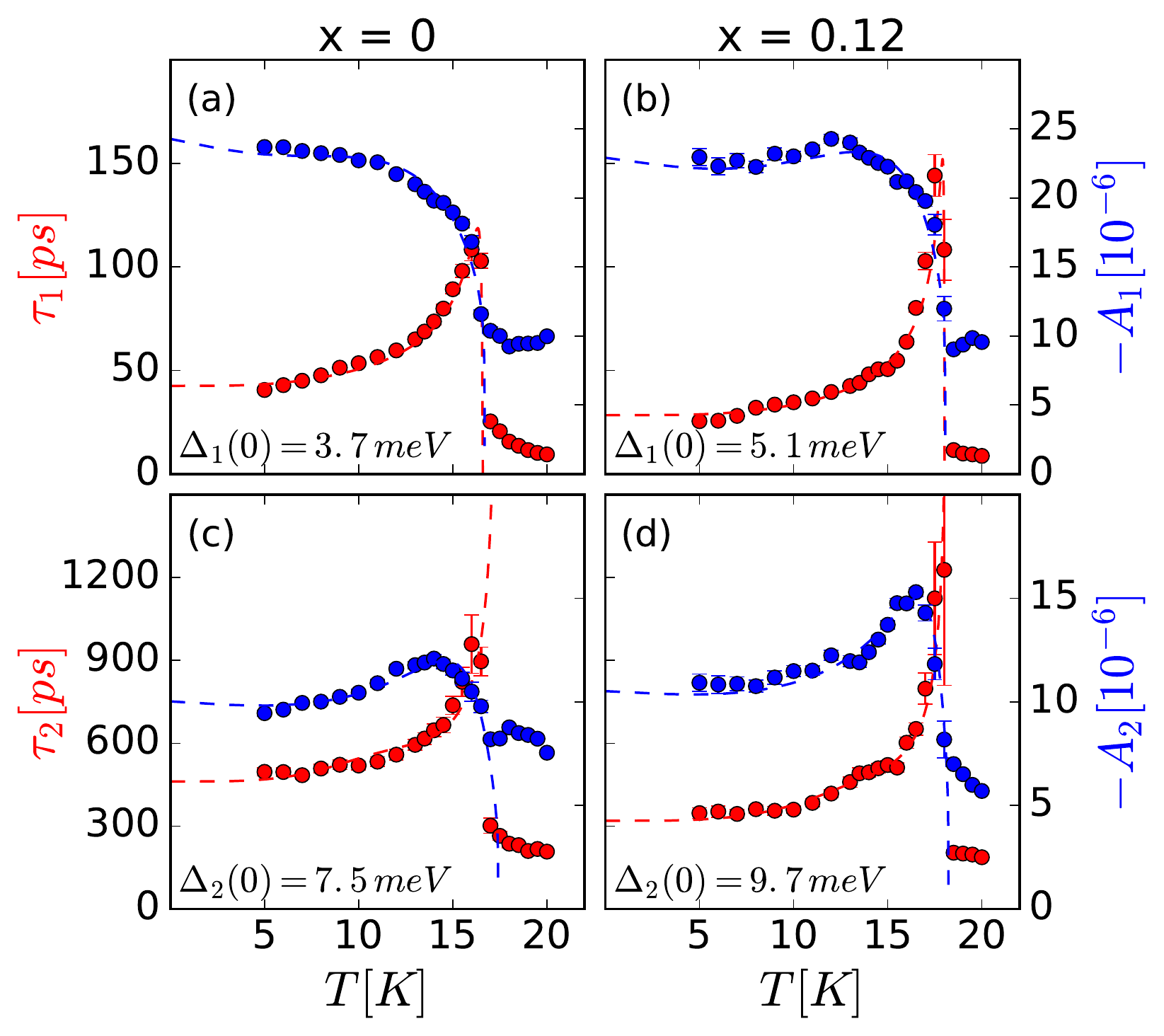}
\caption{$T$ dependence of fit parameters below $T_{0}$. $A_{1}$ and $\tau_{1}$ vs. $T$ for (a) $x$ = 0 and (b) $x$ = 0.12. $A_{2}$ and $\tau_{2}$ vs. $T$ for (c) $x$ = 0 and (d) $x$ = 0.12. Blue and red dashed lines are fits to Eqn. \ref{eq:2} and Eqn. \ref{eq:3}, respectively. Gap energies extracted by these fits are displayed. \label{fig:2}}
\end{figure}

Fig. \ref{fig:2} shows the parameters extracted from fits to the raw data using Eqn. \ref{eq:1} below 20 K. All time constants diverge approaching $T_{0}$ from below and jump to lower values in the PM phase. This divergence is characteristic of a bottleneck associated with a $T$ dependent gap where the limiting step is the anharmonic decay of HEBs \cite{Demsar2001, Lobo2005}. To analyze the $T$ dependence of the fit parameters, we use a bottleneck model due to Kabanov \textit{et al.} \cite{Kabanov1999}:
\begin{equation} \label{eq:2}
A(T) = \frac{F/(\Delta (T)+k_{B}T/2)}{1+\gamma\sqrt{\frac{2k_{B}T}{\pi\Delta (T)}}exp[-\Delta (T)/k_{B}T]}
\end{equation}
\begin{equation} \label{eq:3}
\frac{1}{\tau(T)}=\frac{K\Delta(T)^{2}}{ln(1/\{c\Delta(0)^{2}+exp[-\Delta(T)/k_{B}T]\})}
\end{equation}
In Eqn. \ref{eq:2} and Eqn. \ref{eq:3}, $F \propto \mathcal{E}_{I}$, $\gamma = 2\nu/N(0)\hbar\Omega_{c}$, $K =12\Gamma_{\omega}/\hbar\omega^{2} $, $c = \mathcal{E}_{I}/2N(0)$. $\mathcal{E}_{I}$ is the photoexcited energy density per unit cell, $\nu$ is the number of modes per unit cell, and $N(0)$ is the electronic DOS at $E_{F}$. $\Omega_{c}$, $\Gamma_{\omega}$, and $\omega$ are the cutoff frequency, linewidth, and frequency of the modes, respectively. Eqn. \ref{eq:2} and Eqn. \ref{eq:3} can be derived from the RT model in the strong bottleneck regime. The $T$ dependence of the gaps is modeled with a generic BCS form $\Delta(T) = \Delta(0)tanh(1.74\sqrt(T/T_{0}-1))$ \cite{Lobo2015, Bachar2016}. We treat the zero-$T$ gap $\Delta(0)$ and the transition temperature $T_{0}$ as shared parameters and fit to Eqn. \ref{eq:2} and Eqn. \ref{eq:3} simultaneously, strongly constraining the extracted values of $\Delta(0)$.

Fits to the fast component, shown in (a) and (b) of Fig. \ref{fig:2}, yield smaller gap energies than reported with optical techniques, $\Delta_{1}(0)=3.7\pm0.1$ meV for $x$ = 0 and $\Delta_{1}(0)=5.1\pm0.2$ meV for $x$ = 0.12. An indirect gap is a possibility, since infrared and Raman spectroscopy only probe direct gaps. The energies roughly agree with the energies of the magnetic excitation at $Q_{1}$, which has been interpreted as an interband transition across an indirect hybridization gap \cite{Butch2015, Butch2016}. A hybridization gap bottleneck arises naturally from this interpretation.

Fits to the slow component, shown in (c) and (d), return values of $\Delta_{2}(0)=7.5\pm0.5$ meV and $\Delta_{2}(0)=9.7\pm1.1$ meV for $x$ = 0 and $x$ = 0.12, respectively. These values are consistent with measurements of the charge gaps in the HO and LMAFM phases of Fe-substituted samples \cite{Kanchanavatee2011, Das2015, Hall2015}. The value for $x$ = 0 also agrees with the HO gap from Raman spectroscopy \cite{Buhot2014, Kung2015}, so we interpret the slow component as a bottleneck involving a direct gap between a localized, occupied $f$-state and a light conduction band that crosses the $E_{F}$ as in \cite{Kung2015}.

The gap energies extracted by the fits shown in Fig. \ref{fig:2} correspond to the literature values for both phases. This excellent agreement supports the description of the QP relaxation dynamics in terms of bottlenecks using Eqn. \ref{eq:2} and Eqn. \ref{eq:3} and demonstrates the sensitivity of our technique to the low energy electronic structure of URu$_{2}$Si$_{2}$. Clearly, we can distinguish between the QP relaxation dynamics in the HO and LMAFM phases, even though the gaps of the two phases differ by only a few meV. 

\begin{figure} \centering \includegraphics[width=3.375 in,clip]{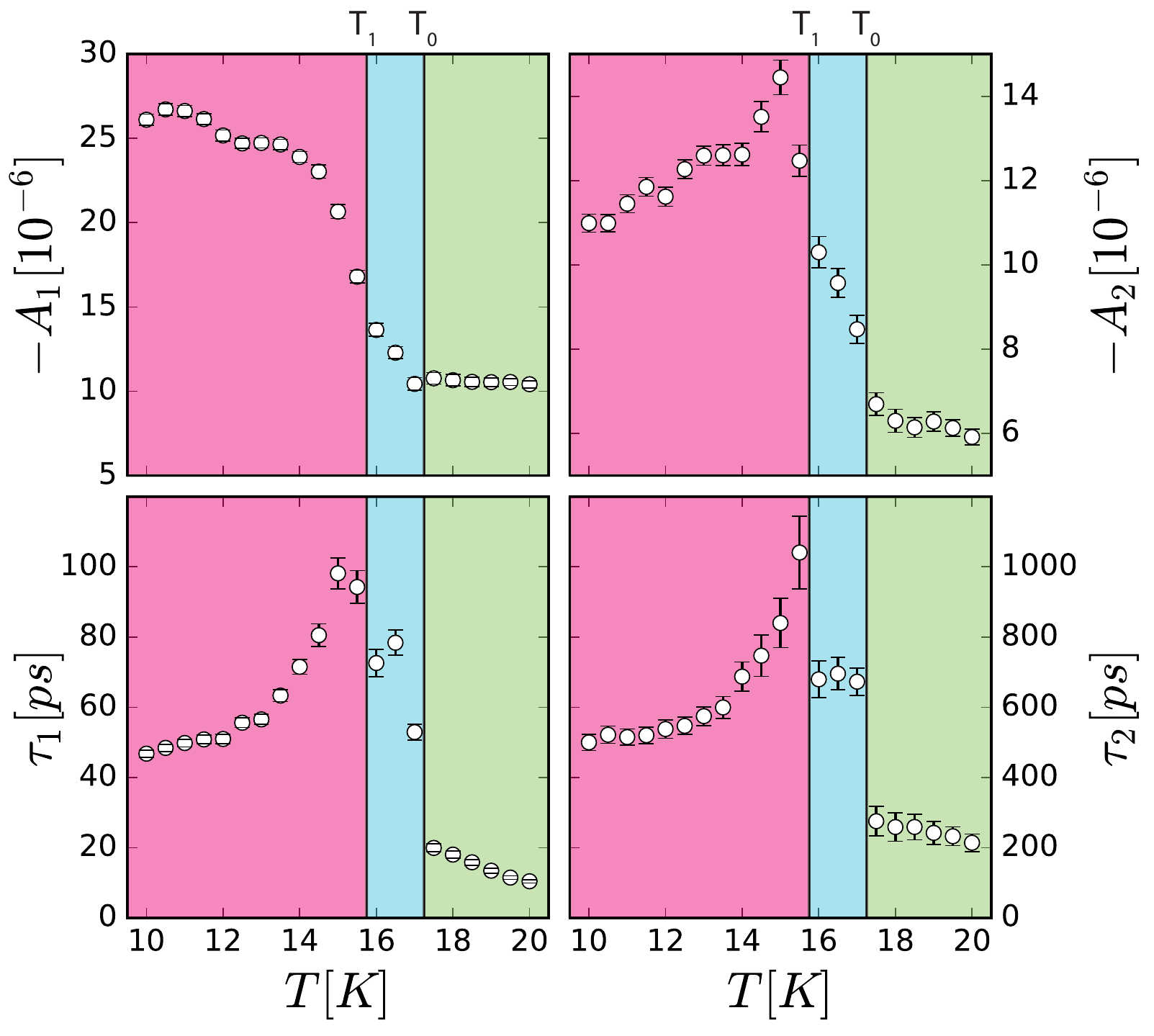}
\caption{$T$ dependence of fit parameters below $T_{0}$ for $x$ = 0.1. The data from the proposed LMAFM, HO, and PM phases are highlighted in red, blue, and green, respectively. \label{fig:3}}
\end{figure}

Armed with an understanding of the dynamics for $x$ = 0  and $x$ = 0.12, we turn to the fit parameters for $x$ = 0.1, shown in Fig. \ref{fig:3}. Anomalous $T$ dependence is observed between 17.5 K and 15.5 K. Both time constants jump twice: once between 17.5 and 17 K at $T_{0}$, and again at a second temperature $T_{1}$ between 16 and 15.5 K. In contrast, abrupt changes in relaxation times occur only once, at $T_{0}$, for both $x$ = 0 and $x$ = 0.12. Additionally, the rise in amplitudes below $T_{0}$ occurs more gradually for $x$ = 0.1 than for $x$ = 0 or $x$ = 0.12, with a discontinuity in slope at $T_{1}$. These observations are reminiscent of thermal expansion measurements \cite{Ran2016}, where the two phase transitions observed for $x$ = 0.08 and $x$ = 0.1 were interpreted as a second order PM to HO transition and a first order HO to LMAFM transition.

\begin{figure} \centering \includegraphics[width=3.375 in,clip]{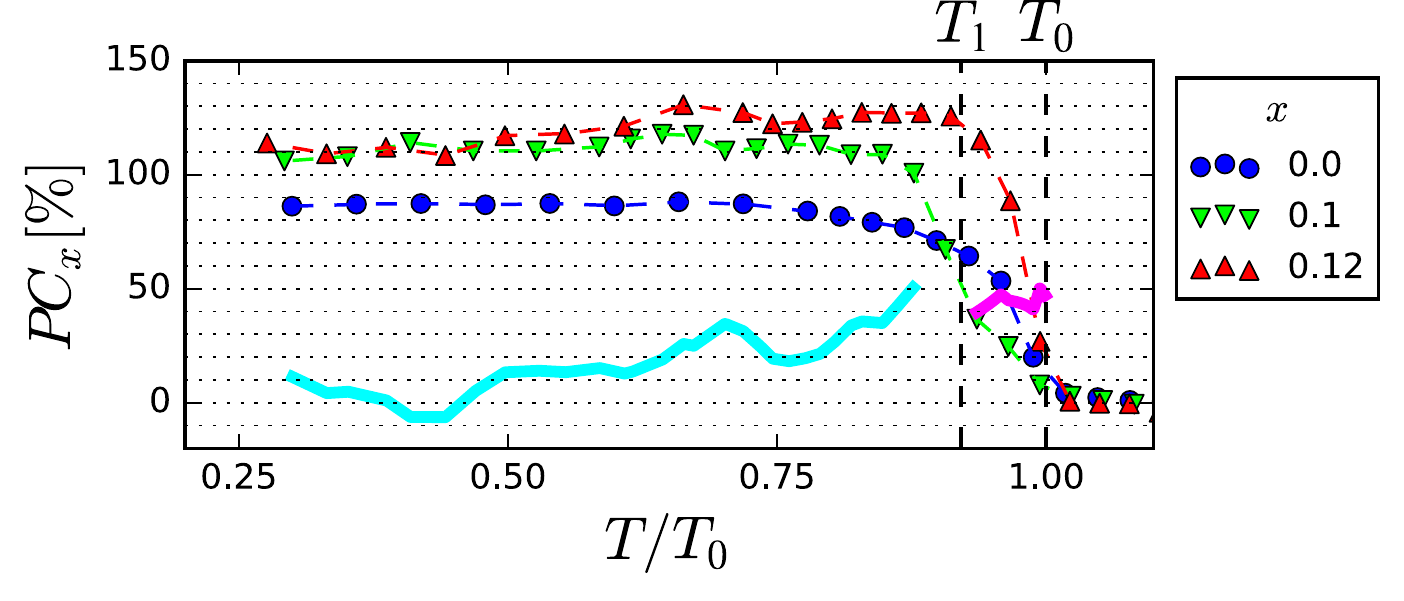}
\caption{Percent change in OPOP signal amplitude $PC_{x}$ vs. reduced temperature $T/T_{0}$ for $x$ = 0, $x$ = 0.1, and $x$ = 0.12. Dashed colored lines are interpolations between adjacent data points. Dashed vertical lines are estimates for $T_{0}$ and $T_{1}$. The solid cyan line is an estimated HO volume fraction $V_{HO}$ below $T_{1}$ \cite{Supp}. The solid magenta line is an estimated PM volume fraction $V_{PM}$ between $T_{0}$ and $T_{1}$ \cite{Supp} based on the possible phase coexistence between HO and PM. $V_{HO}$ and $V_{PM}$ are listed as percentages to facilitate plotting on the same axes.
\label{fig:4}}
\end{figure}

One particularly striking feature of the data is the abrupt increase in signal amplitude observed at the HO and LMAFM transitions. We perform a model-independent comparison of the low $T$ behavior of all three samples by considering the percentage change in the signal amplitude, $PC_{x}(T)$ \cite{Supp}, \textit{with respect to the PM phase}, in which the dynamics are independent of $x$ \cite{Supp}. These quantities are plotted for each sample as a function of reduced temperature $T/T_{0}$ in Fig. \ref{fig:4}. For $x$ = 0, the amplitude nearly doubles as the sample cools from the PM phase to the HO phase (corresponding to a percentage change of nearly 100$\%$). The increase is even greater for LMAFM phase in $x$ = 0.12.

The well documented coexistence of HO and LMAFM domains in the parent compound \cite{Matsuda2001, Amitsuka2003} originates from inhomogeneous strain due to defects \cite{Yokoyama2005}, and is thus likely to be enhanced around Fe sites, as in the case of Rh substituted samples \cite{Baek2010}. This effect likely plays a much larger role for $x$ = 0.1, given its proximity to the HO/LMAFM phase boundary, than for either $x$ = 0 and $x$ = 0.12. Therefore, in order to study inhomogeneity in $x$ = 0.1, we assume that $x$ = 0 and $x$ = 0.12 represent comparatively pure HO and LMAFM phases, respectively.

Below $T_{1}$, the data for $x$ = 0.1 matches expectations. Since the amplitude in Eq. \ref{eq:2} depends only on band structure and lattice parameters \cite{Kabanov1999}, the overlap between $PC_{0.1}(T)$ and $PC_{0.12}(T)$ suggests that $x$ = 0.1 and $x$ = 0.12 have the same phase composition at low $T$. Closer to $T_{1}$, $PC_{0}(T) < PC_{0.1}(T) < PC_{0.12}(T)$. This is expected behavior for coexisting HO and LMAFM domains, with a HO volume fraction $V_{HO}$ \cite{Supp} that decreases with $T$. As seen in Fig. \ref{fig:4}, $V_{HO}$ decreases roughly linearly, from $V_{HO} \approx$ 0.5 just below $T_{1}$ to $V_{HO} \approx$ 0 at low $T$. In contrast, it is difficult to describe the data for $x$ = 0.1 between $T_{0}$ and $T_{1}$ in terms of coexistence between HO and LMAFM. Here, the expected phase composition is primarily HO with a small LMAFM volume fraction. However, $PC_{0.1}(T)$ is less than expected for both pure HO and LMAFM.

We speculate that the anomalously small signal amplitude arises from phase coexistence of PM and HO. For example, if we assume that the signal for $x$ = 0.1 contains contributions from HO and PM domains, we obtain a PM volume fraction $V_{PM} \approx$ 0.5 \cite{Supp}, which is strikingly close to $V_{HO}$ just below $T_{1}$. There are several reasonable explanations for this unusual possibility. Perhaps the PM to HO transition in this sample is driven weakly first order by proximity to the bicritical point in the phase diagram or by disorder from Fe substitution. On the other hand, there is evidence for a weakly first order PM to HO transition in the parent compound \cite{Tonegawa2014}, and a first order PM to HO transition was predicted in a recent theoretical study \cite{Shen2018}. It is also possible that this is a nonequilibrium effect, similar to the coexistence between superconducting and normal state domains observed in photoexcited superconductors \cite{Gianetti2009,Coslovich2011,Matsunaga2012}. Each of these outcomes points to exotic and novel physics in $f$-electron systems, meriting future studies to replicate this observation and to clarify its origin, if confirmed.

\begin{figure} \centering \includegraphics[width=3.375 in,clip]{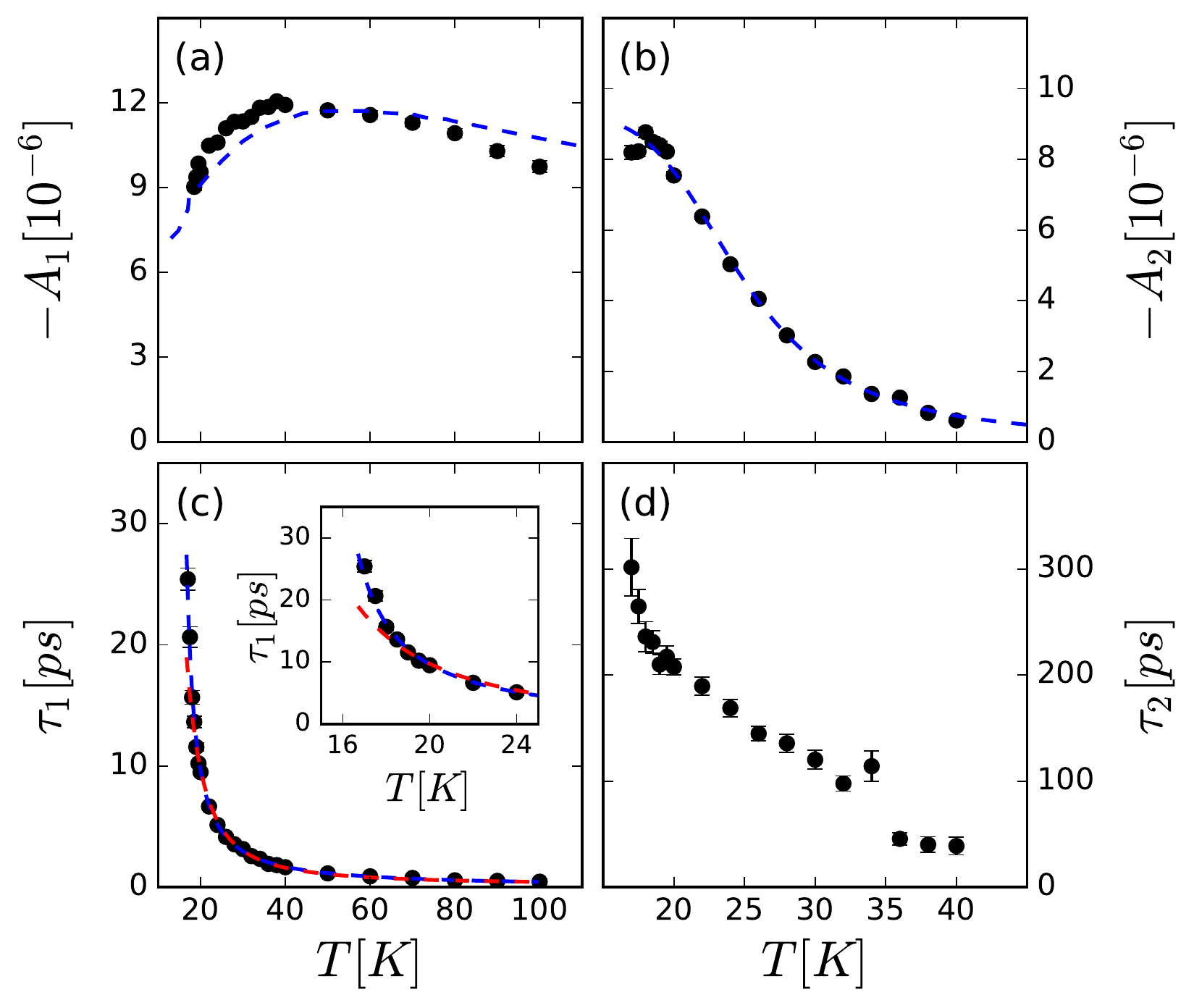}
\caption{T dependence of fit parameters in the PM phase for $x$ = 0. (a) $A_{1}$ vs. T. The blue dashed line is the $c$-axis magnetic susceptibility from \cite{Maple1986}, scaled for comparison. (b) $A_{2}$ vs. T. The blue dashed line is a fit to Eq. \ref{eq:2} with a $T$ independent gap \cite{Kabanov1999}. (c) $\tau_{1}$ vs. T. The blue and red dashed lines are fits to a power law and a bottleneck model, respectively, as described in the text. The inset shows the same quantities plotted over a narrower $T$ range, from 15 K to 25 K. (d) $\tau_{2}$ vs. T. \label{fig:5}}
\end{figure}

The data in the PM phase is also informative. We first discuss the $T$ dependence of $A_{2}$, shown in Fig. \ref{fig:5}(b). In a previous OPOP study on the parent compound \cite{Liu2011}, this component was interpreted as evidence of a HO pseudogap (HOPG) \cite{Levallois2011, Haraldsen2011, Shirer2013}. The dynamics we observe are nearly independent of \textit{x} in the PM phase \cite{Supp}. This observation is likely incompatible with a HOPG that arises from competition between HO and LMAFM phases, but does not rule out scenarios where the HOPG originates from fluctuations of a shared HO-LMAFM order parameter above $T_{0}$ \cite{Kung2016, Niklowitz2014}. On the other hand, a fit of $A_{2}$ to a \textit{T}-independent form of Eq. \ref{eq:2} between $T_{0}$ and 40 K, shown in Fig. \ref{fig:5}(b), is consistent with a bottleneck associated with a gap of 13.8$\pm$0.5 meV. This is likely the correct interpretation of this component, since a hybridization gap of similar energy has been observed in the PM phase with a number of techniques \cite{Lobo2015,Bachar2016,Levallois2011,Park2012}.

The fast process $A_{1}$ and $\tau_{1}$ is more difficult to interpret. As shown in Fig. \ref{fig:5}, $A_{1}$ peaks near 40 K and decreases upon approaching $T_{0}$. This is not the expected behavior from a bottleneck. At this $F$, we assume that the concentration of photoexcited QPs is much less than the concentration of thermal QPs $n_{S}\ll n_{T}$ and approximate $\tau^{-1}(T)=C[n_{S}+n_{T}]\approx Cn_{T}$, with $n_{T}=T^{1/2}e^{-\Delta/kT}$. A fit to this equation, shown in red in Fig. \ref{fig:5}(c), returns a gap value of $\Delta$=4.9$\pm$0.1 meV. This value does not match any charge gap reported in the literature above $T_{0}$.

On the other hand, the resemblance between $A_{1}$ and the $c$-axis magnetic susceptibility, highlighted in Fig. \ref{fig:5}(a), indicates that the fast process may have a magnetic origin. A power law fit of the form $\tau_{1}(T)\propto((T-T_{0})/T_{0})^{-k}$ \cite{Zhu2018}, shown in Fig. \ref{fig:5}(c), reproduces the $T$ dependence of $\tau_{1}$, particularly the quasi-divergence near $T_{0}$, with $k=1.14\pm0.05$ and $T_{0}=14.5\pm0.4$ K. The slight disagreement between the nominal and extracted values of $T_{0}$ is likely due to pump induced heating that limits the accuracy of transition temperatures and critical exponents measured with this technique. Nonetheless, the value for the scaling exponent $k$ is close to the exponent describing critical slowing down in the 3D Ising model, $\nu z=1.28\pm0.03$ \cite{Wansleben1987, Pelissetto2002}, where $\nu$ is the critical exponent of correlation length and $z$ is the dynamical critical exponent \cite{Hohenberg1977}. The 3D Ising model is a good starting point to describe magnetic fluctuations (MFs) in the PM phase given the notable Ising anisotropy in URu$_{2}$Si$_{2}$ \cite{Ramirez1992}. We conclude that the fast process tracks a relaxation channel for photoexcited QPs involving scattering with MFs that slows as the MFs become critical. The literature supports this interpretation. Strong MFs are present at both $Q_{0}$ \cite{Bourdarot2010} and $Q_{1}$ \cite{Wiebe2007} in this $T$ range, and nearly critical behavior of MFs at $Q_{0}$ has been reported \cite{Niklowitz2014}. The THz frequency scattering of carriers by critical MFs can also explain the non-Fermi liquid behavior observed in the PM state \cite{Nagel2012}.

To conclude, our measurements of QP relaxation dynamics in the URu$_{2-x}$Fe$_{x}$Si$_{2}$ single crystals reveal several new insights. The dynamics in the PM phase, which are nearly independent of $x$, highlight the presence of a hybridization gap as well as the influence of strong interactions between QPs and critical MFs. Below $T_{0}$, the dynamics in the HO and LMAFM phases reflect known differences in the low energy electronic structure. As in past measurements \cite{Ran2016}, we observe a second phase transition in a sample of intermediate Fe substituent concentration $x$ = 0.1. In addition to a low $T$ LMAFM phase, there is a distinct intermediate $T$ HO phase. The anomalous data in this phase suggests the unexpected possibility of coexisting  HO and PM. Our study lays the groundwork for future experiments on the URu$_{2-x}$Fe$_{x}$Si$_{2}$ system to understand HO, its relationship to LMAFM, and novel forms of order in $f$-electron systems more generally. 

\begin{acknowledgments}
This research was supported by the U.S. National Science Foundation under Grant No. DMR-1810310 (ultrafast optical pump-probe spectroscopy measurements and materials characterization), the U.S. Department of Energy, Office of Science, Office of Basic Energy Sciences under Award Number DE-FG02-04ER46105 (crystal growth), and ARO W911NF-16-1-0361 (ultrafast instrumentation development).
\end{acknowledgments}

\bibliography{Refs}

\end{document}

\title{Supplementary Materials: Quasiparticle Relaxation Dynamics in URu$_{2-x}$Fe$_{x}$Si$_{2}$ Single Crystals}
\author{Peter Kissin}
\affiliation{Department of Physics, University of California San Diego, 9500 Gilman Drive, La Jolla, California 92093, USA}

\author{Sheng Ran}
\altaffiliation[Present Addresses: ]{Center for Nanophysics and Advanced Materials, Department of Physics, University of Maryland, College Park,
MD 20742; NIST Center for Neutron Research, National Institute
of Standards and Technology, 100 Bureau Drive, Gaithersburg, MD
20899.}
\affiliation{Department of Physics, University of California San Diego, 9500 Gilman Drive, La Jolla, California 92093, USA}
\affiliation{Center for Advanced Nanoscience, University of California San Diego, La Jolla, California 92093, USA}

\author{Dylan Lovinger}
\affiliation{Department of Physics, University of California San Diego, 9500 Gilman Drive, La Jolla, California 92093, USA}

\author{Verner K. Thorsm\o lle}
\affiliation{Department of Physics, University of California San Diego, 9500 Gilman Drive, La Jolla, California 92093, USA}

\author{Noravee Kanchanavatee}
\altaffiliation[Present Address: ]{Department of Physics, Chulalongkorn University, Pathumwan, 10330, Thailand.}
\affiliation{Department of Physics, University of California San Diego, 9500 Gilman Drive, La Jolla, California 92093, USA}
\affiliation{Center for Advanced Nanoscience, University of California San Diego, La Jolla, California 92093, USA}

\author{Kevin Huang}
\altaffiliation[Present Address: ]{National High Magnetic Field Laboratory, Florida State University, Tallahassee, FL 32313.}
\affiliation{Center for Advanced Nanoscience, University of California San Diego, La Jolla, California 92093, USA}
\affiliation{Materials Science and Engineering Program, University of California San Diego, 9500 Gilman Drive, La Jolla, California 92093, USA}

\author{M. Brian Maple}
\affiliation{Department of Physics, University of California San Diego, 9500 Gilman Drive, La Jolla, California 92093, USA}
\affiliation{Center for Advanced Nanoscience, University of California San Diego, La Jolla, California 92093, USA}

\author{Richard D. Averitt}
\email[Corresponding Author:]{raveritt@ucsd.edu}
\affiliation{Department of Physics, University of California San Diego, 9500 Gilman Drive, La Jolla, California 92093, USA}
\date{\today}

\maketitle

\section{S1: Experimental Setup}
\begin{figure} \centering \includegraphics[width=6.75 in,clip]{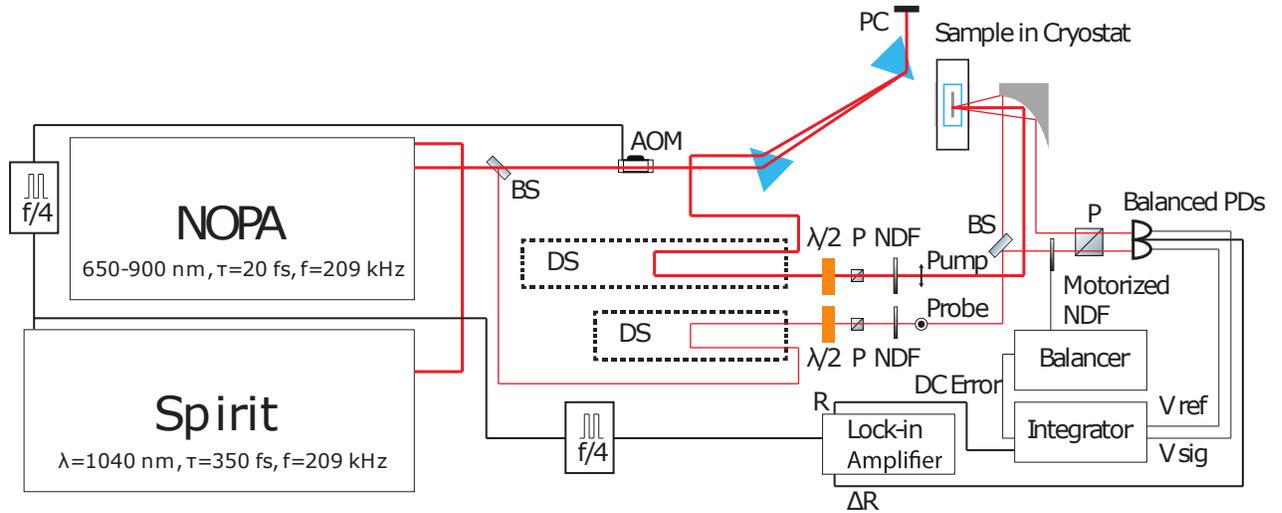}
\caption{Schematic of our laser system and experiment. BS = beamsplitter, AOM = acousto-optic modulator, PC = prism compressor, DS = delay stage, $\lambda/2$ = half waveplate. P = polarizer, NDF = continuously variable neutral density filter}.
\end{figure}
Our laser system consists of an amplified Yb fiber laser, producing 350 fs pulses centered at 1040 nm. 4 W of 1040 nm is frequency doubled to 520 nm and used to pump a non-collinear optical parametric amplifier with two amplification stages. This setup produces several hundred mW of power depending on signal wavelength, tunable from 650-900 nm, with a bandwidth-limited pulse duration of less than 20 fs. We use an AOM made of quartz (for reduced chromatic dispersion compared to the more common TeO$_{2}$) to modulate the pump beam for lock-in detection of the signal. This is synchronized with a subharmonic of the laser repetition rate, so that every pump pulse is either diffracted or blocked. Achromatic focusing on to the sample is accomplished using an off-axis parabolic mirror. 

Balanced photodiodes are used to enhance the dynamic range of detection and to partially remove shot-to-shot fluctuations of the laser intensity from the signal. We find that balanced detection increases the smallest signal we can resolve by slightly less than an order of magnitude. Small long term drifts in the balance arise from temperature dependent changes in reflectivity, changes in the position of beam on sample, formation of ice on the sample surface, etc. These reduce the effectiveness of balanced detection. In order to correct for this drift, the signal from each photodiode is integrated, producing a DC signal. The two DC signals are subtracted, producing a difference signal that drives a servo motor. The servo motor adjusts the position of an ND filter placed in the reference beam path in order to set the intensity of the reference beam equal to that of the signal beam. The DC signal from the sample photodiode is sent to the auxiliary channel of lock-in amplifier for continuous measurement of the reflectivity R.

\section{S2: Analysis of Laser Heating}
The use of an intermediate repetition rate amplifier has several advantages over the more common 80 MHz oscillator for low temperature OPOP measurements. For an 80 MHz system, cumulative heating due to the average optical power incident upon the illuminated spot is a large source of temperature uncertainty even at low fluences. In contrast, the repetition rate of 209 kHz is more than two orders of magnitude lower, allowing us to study a wider range of pump excitation densities without heating the sample at low temperatures. This effect likely accounts for the qualitative difference in fluence dependence between our data and the previous OPOP data on the parent compound reported by Liu $\textit{et al.}$ \cite{Liu2011}. 

At the primary pump fluence of 0.5 $\mu$J/cm$^{2}$ used in this study, analysis of the specific heat reported in \cite{Maple1986} bounds the temperature change in the sample to less than 4 K at a sample temperature of 5 K and by less than 0.5 K just below $T_{0}\approx16.5K$ in the parent compound. The number of nonequilibrium e-h pairs created is an alternate way to assess the strength of photoexcitation. Assuming the 7 meV gap feature in optical conductivity at ~5 K \cite{Bonn1988, Guo2012, Bachar2016, Lobo2015, Hall2012} roughly corresponds to the energy of the QP excitation to which our OPOP measurements are sensitive, we estimate that 0.002 e-h pairs/U atom are e-h pairs are excited by the pump pulse at this fluence. This is an order of magnitude lower than the thermal carrier density inside the HO phase of 0.02 e-h pairs/U atom \cite{Kasahara2007}, ensuring that our measurements are performed in the weak photoexcitation regime.
\section{S3: Additional Raw Data and Full Temperature Dependence of Fit Parameters}
\begin{figure}\includegraphics[width=3.375 in,clip]{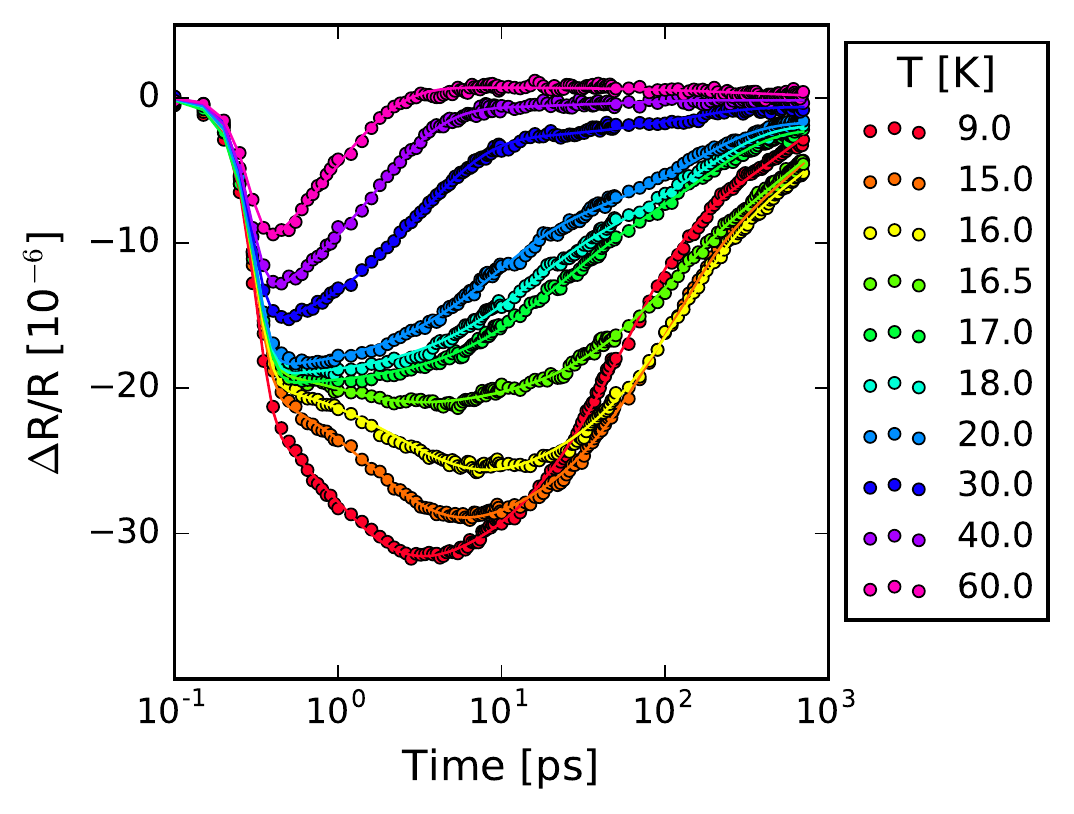}
\includegraphics[width=3.375 in,clip]{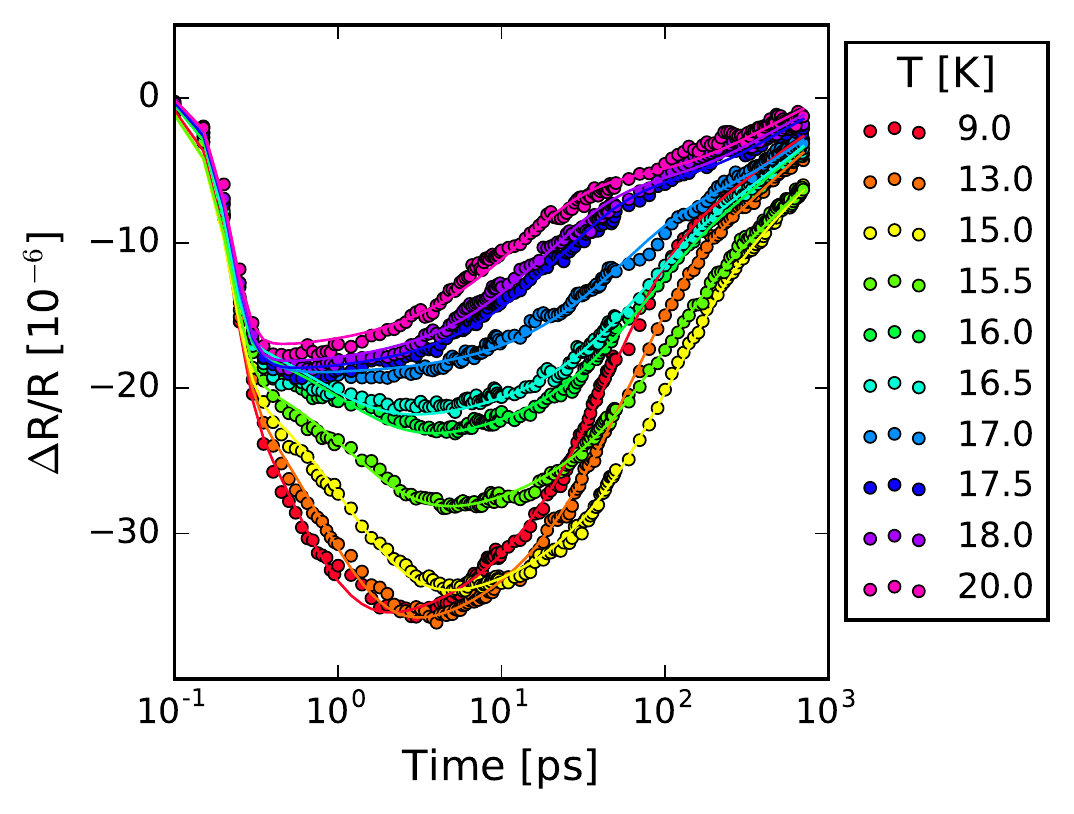}
\includegraphics[width=3.375 in,clip]{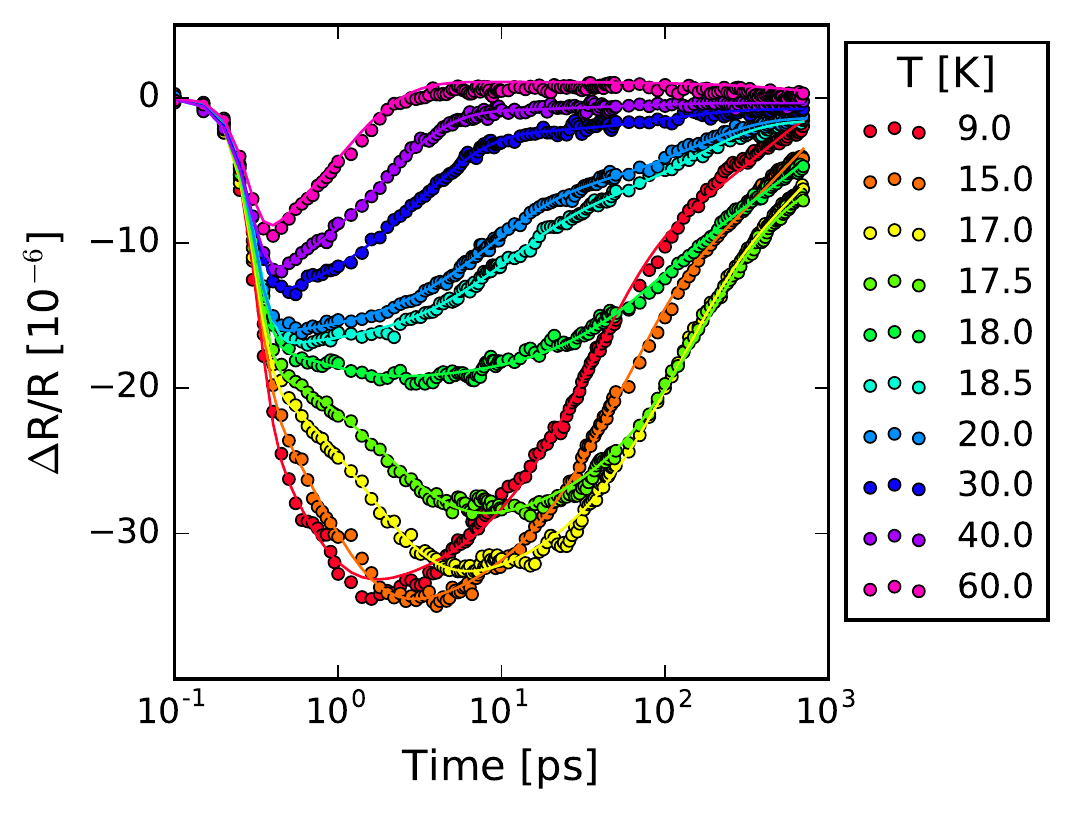}
\caption{Raw data for Fe substituted samples. Top: x=0. Middle: x=0.1. Bottom: x=0.12.}.
\end{figure}
\begin{figure}\includegraphics[width=6.75 in,clip]{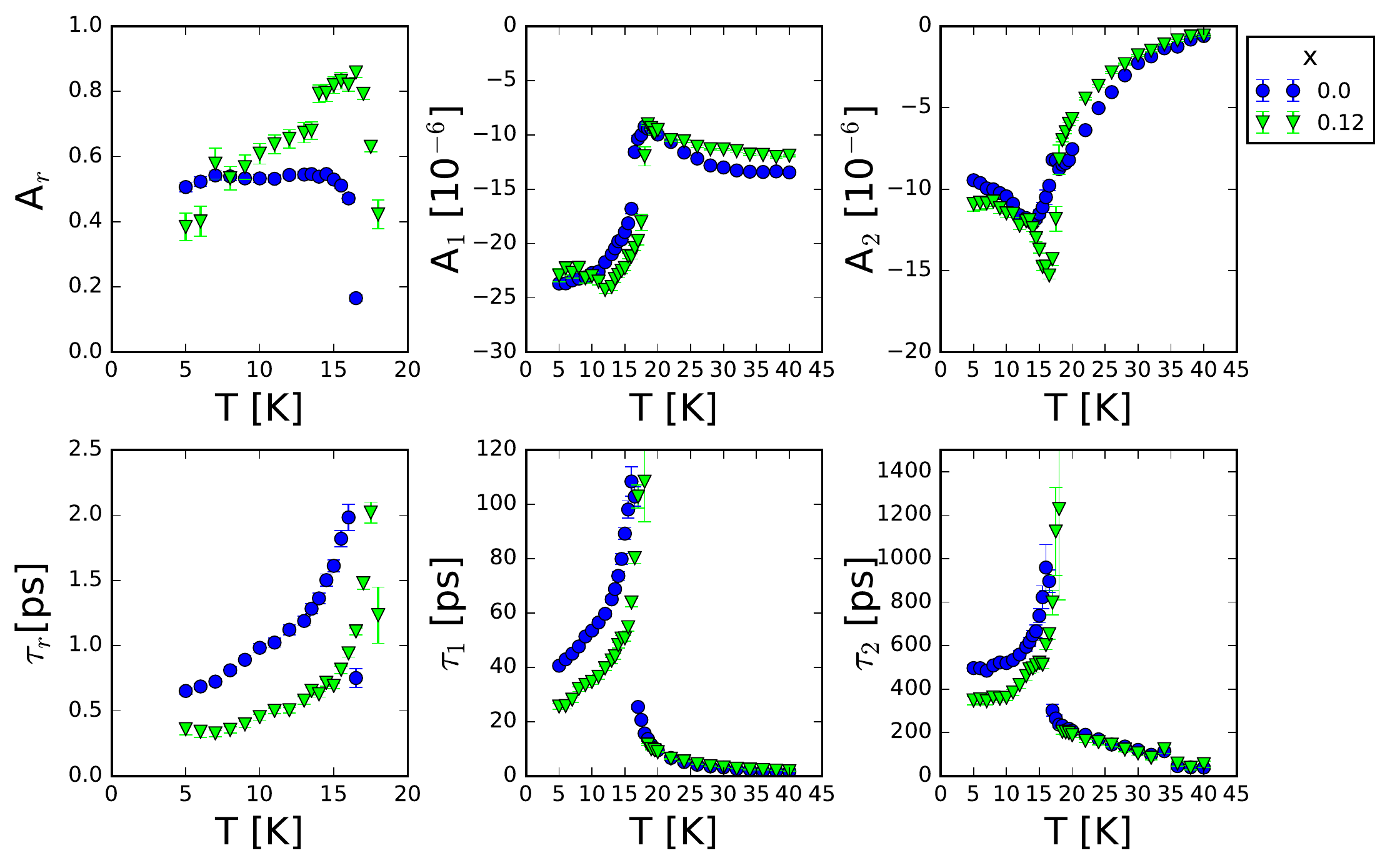}
\caption{Full temperature dependence of fit parameters for x=0, x=0.12.}.
\end{figure}
Additional raw data for Fe substituted samples is shown in Fig. S2. The full temperature dependence of the fit parameters for x=0 and x=0.12, including the rise dynamics, is shown in Fig. S3.
\section{IV: Origin of the Slow Rise Dynamics}
The temperature dependence of the anomalous rise dynamics observed below $T_{0}$ merits comment. Typically, the rise dynamics occurring within the first $\approx$100 fs of photoexcitation are a result of QP multiplication triggered by the scattering of hot photoexcited carriers as they relax to the gap edges \cite{Demsar2006_2}. However, this description is inadequate for rise dynamics occurring on a longer timescale. Similar slow rise dynamics have been observed in low fluence pump probe experiments on superconductors \cite{Demsar2003} and heavy fermion compounds \cite{Demsar2003_2}. Slow rise dynamics can emerge from the RT model in the case where hot electrons relax to $E_{F}$ preferentially through electron-boson scattering, creating an excess population of high energy bosons \cite{Kabanov2005}. While we cannot rule out such a scenario for URu$_{2}$Si$_{2}$, the fact that the slow rise dynamics are fluence-independent, onset abruptly at $T_{0}$ without significantly altering the initial fast rise time, and that $\tau_{r}$ diverges upon warming to $T_{0}$ point to an origin more directly related to the low energy bottleneck. One possible scenario is the transfer of electronic energy from the subsystem of quasiparticles that have relaxed to the edge of the hybridization gap that exists already above $T_{0}$ to another subsystem of quasiparticles more directly related to the HO/LMAFM charge gaps. Time resolved photoemission experiments with extremely high energy resolution may help to clarify the origins of the slow rise in URu$_{2}$Si$_{2}$.
\section{V: Fluence Dependence}
\begin{figure}\includegraphics[width=6.75 in,clip]{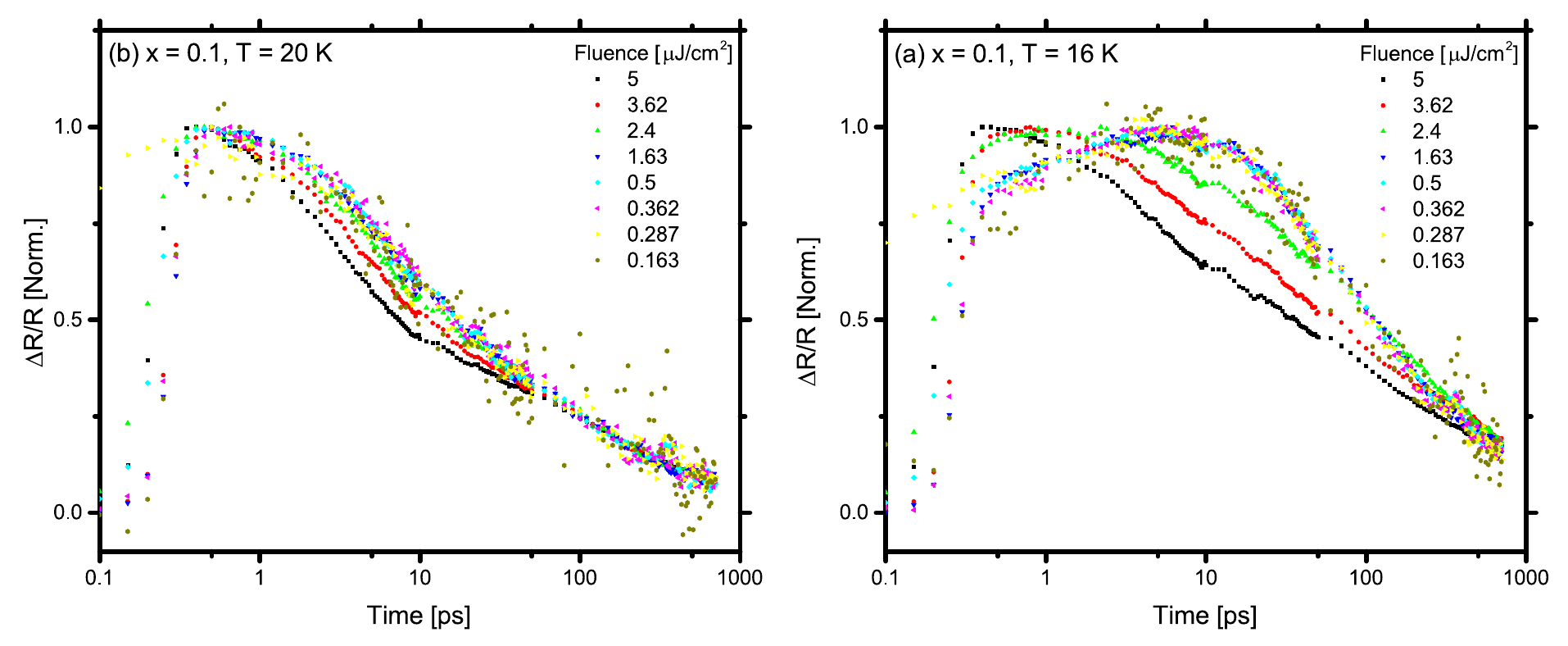}
\caption{Fluence dependence of the data (a) above $T_{0}$, and (b) below $T_{0}$.}.
\end{figure}

Fig. S4 shows the fluence dependence of the data covering more than an order of magnitude, roughly centered on 0.5 $\mu$J/cm$^{2}$, above and below $T_{0}$, for x = 0.1, which should be representative of all samples. For both phases, when the data is normalized, all the of the curves overlap, showing that the data is independent of pump fluence in this range of fluences. At higher fluences, above roughly 2 $\mu$J/cm$^{2}$, some fluence dependence is observed. This fluence corresponds to the crossover where the density of photoexcited quasiparticles approaches the density of thermal quasiparticles, meaning that the weak photoexcitation limit no longer holds.

\section{S5: Estimating Percentage Change and HO Volume Fraction}
For our analysis of heterogeneity, we consider the quantity \\
$PC_{x}(T) = (A_{x}(T)-(A_{x}(20 K))/(A_{x}(20 K)$, where $A_{x}(T)$ denotes the raw signal amplitude for sample $x$ at temperature $T$. Since the PM component of the signal is independent $x$ and varies relatively slowly with $T$, $PC_{x}(T)$ has the approximate form of a percentage change of the signal amplitude due to the emergence of HO or LMAFM order for each sample.

Below $T_{1}$, $PC_{0.1}(T)$ is larger than in the signal in the pure HO phase $PC_{0}(T<T_{1})$ but smaller than in the the signal in the pure LMAFM phase $PC_{0.12}(T<T_{1})$. So, we assume that the signal for $x$ = 0.1 in this temperature range consists of HO regions with volume fraction $V_{HO}$ and LMAFM regions with volume fraction $1-V_{HO}$. Thus, $PC_{0.1}(T<T_{1}) = (1-V_{HO})PC_{0.12}(T<T_{1})+V_{HO}PC_{0}(T<T_{1})$, which we write as $V_{HO}=(PC_{0.12}-PC_{0.1})/(PC_{0.12}-PC_{0})$.

Between $T_{0}$ and $T_{1}$, $PC_{0.1}(T)$ is larger than in the signal in the PM phase ($PC_{PM}\approx1$ by definition) but smaller than in the the signal in the pure HO phase $PC_{0}(T<T_{0})$ and much smaller than in the signal in the pure LMAFM phase $PC_{0.12}(T<T_{0})$. So, we assume that the signal for $x$ = 0.1 in this temperature range consists of PM regions with volume fraction $V_{PM}$ and HO regions with volume fraction $1-V_{PM}$. Thus, $PC_{0.1}(T_{1}<T<T_{0}) = (1-V_{PM})PC_{0}(T_{1}<T<T_{0})+V_{PM}$, which we write as $V_{PM}=(PC_{0}-PC_{0.1})/(PC_{0}-1)$.

\section{S6: Effect of Fe Substitution in the Paramagnetic Phase}
\begin{figure}\includegraphics[width=3.375 in,clip]{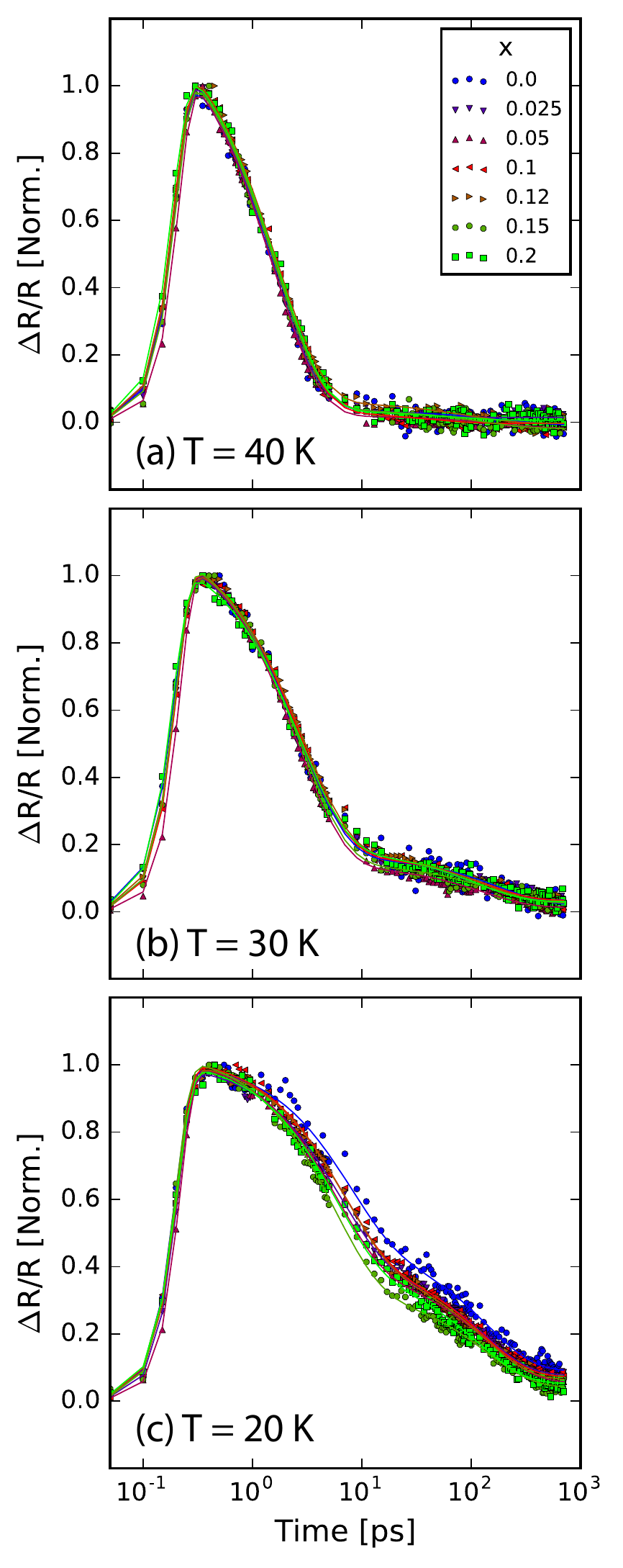}
\caption{Normalized raw data for samples of various Fe-substitution in the paramagnetic phase. (a) $T$ = 40K, (a) $T$ = 30K, (c) $T$ = 20K.}.
\end{figure}
As shown in Fig. S5, while the relaxation dynamics are independent of x at 30 K, a subtle dependence on x is present at lower temperatures and enhanced upon approaching $T_{0}$. This observation is suggestive of hidden-order pseudogap physics, and there is a possible trend towards faster relaxation with increasing x. However, the effect is not systematic and can be explained equally well by a number simpler scenarios. For example, due to the quasidivergence of $\tau_{1}$ upon approaching $T_{0}$, small variations in the actual temperature of samples due thermal contact with the cold finger will produce large changes in the time dependence of the dynamics. Uncertainties in temperature as high as 1 K are likely given the slight disagreement between the nominal values of $T_{0}$ and the experimentally observed values, which would be large enough to produce this effect. Another possibility is that Fe substitution can result in disorder than enhances the anharmonicity of bosons involved in the relaxation bottleneck. In any case, this effect too weak to conclusively assert its origin in pseudogap physics.

\bibliography{Refs}